\documentclass[aps,prl,twocolumn,groupedaddress]{revtex4}
\usepackage{graphicx,latexsym}
\pdfoutput=1

\begin{document}

\title{Phase diagram of a two-component Fermi gas with resonant interactions}

\author{Yong-il Shin}\email{yishin@mit.edu}
\author{Christian H. Schunck}
\author{Andr\'{e} Schirotzek}
\author{Wolfgang Ketterle}

\affiliation{Department of Physics, MIT-Harvard Center for Ultracold
Atoms, and Research Laboratory of Electronics, Massachusetts Institute of
Technology, Cambridge, Massachusetts, 02139, USA}

\date{\today}

\begin{abstract}
The pairing of fermions is at the heart of superconductivity and
superfluidity.  The recent experimental realization of strongly
interacting atomic Fermi gases has opened a new, controllable way to
study novel forms of pairing and superfluidity. A major controversial
issue has been the stability of superfluidity against an imbalance
between the two spin components when the fermions interact resonantly.
Here we present the phase diagram of a spin-polarized Fermi gas of $^6$Li
atoms at unitarity, mapping out the superfluid phase versus temperature
and density imbalance. Using tomographic techniques, we reveal spatial
discontinuities in the spin polarization, the signature of a first-order
superfluid-to-normal phase transition, which disappears at a tricritical
point where the nature of the phase transition changes from first-order
to second-order. At zero temperature, there is a quantum phase transition
from a fully-paired superfluid to a partially-polarized normal gas. These
observations and the implementation of an in situ ideal gas thermometer
provide quantitative tests of theoretical calculations on the stability of
resonant superfluidity.
\end{abstract}


\maketitle

\noindent Superfluidity and superconductivity of fermions are based on
the formation of fermion pairs.  The stability of these pairs determines
the robustness of the superfluid state, and the quest for superconductors
with high critical temperature is a search for systems with strong
pairing mechanisms. Ultracold atomic Fermi gases present a highly
controllable model system for studying strongly interacting
fermions~\cite{GPS07}. Tunable interactions utilizing Feshbach
collisional resonances and control of population or mass imbalance among
the spin components provide unique opportunities to investigate the
stability of pairing~\cite{Chandra62,CLO62,Sar63}, and possibly to search
for new exotic forms of superfluidity~\cite{FF64,LO65}. The case of
unitarity, when the two spin components resonantly interact and the
behavior of the system becomes independent of the nature of the
interactions, has become a benchmark for experimental and theoretical
studies over the last few years. However, there is an ongoing debate
about the stability of resonant superfluidity, reflected in major
discrepancies in predicted transition temperatures for the balanced spin
mixture~\cite{BDM06,BPS06,HRC07}, and an even more dramatic discrepancy
for the critical imbalance of the two spin components, called the
Chandrasekhar-Clogston (CC) limit of
superfluidity~\cite{Chandra62,CLO62}. Recent Quantum Monte-Carlo (QMC)
calculations predicted that superfluidity would be quenched by a density
imbalance around 40\%~\cite{LRG06}, whereas other studies predicted a
critical imbalance above 90\%~\cite{CR05,SR06,GRS06,YD06b,CCH07,PML07}.
Our earlier work~\cite{ZSS06a,ZSS06b,SZS06} suggested the lower limit but
other experiments~\cite{PLK06,PLK06b} were interpreted to be consistent
with the absence of the CC limit. This huge discrepancy reveals that even
qualitative aspects, such as the role of interactions in the normal
phase, are still controversial. The lack of reliable thermometry for
strongly interacting systems limits the full interpretations of
experimental results.

Here we resolve this long standing debate by presenting the phase diagram
of a spin-polarized Fermi gas at unitarity. We observe that the
normal-to-superfluid phase transition changes its nature. At low
temperature, the phase transition occurs with a jump in the spin
polarization as the imbalance increases, which we interpret as a
first-order phase transition. The local spin polarization or local
density imbalance is defined as $\sigma = (n_\uparrow
-n_\downarrow)/(n_\uparrow +n_\downarrow)$, where $\uparrow$ and
$\downarrow$ refer to the two spin components with densities
$n_{\uparrow,\downarrow}$. At high temperature, the phase transition is
smooth and therefore of second-order. The two regimes are connected by a
tricritical point~\cite{Sar63,GRI70} for which we estimate the position as
$(\sigma_{tc}, T_{tc}/T_{F\uparrow}) \approx (0.2, 0.07)$, where $k_B
T_{F\uparrow}= \hbar^2 (6 \pi^2 n_\uparrow)^{2/3}/2m$ is the Fermi energy
of the majority component of density $n_\uparrow$ ($k_B$ is the Boltzmann
constant, $\hbar$ is the Planck constant divided by $2\pi$ and $m$ is the
atomic mass of $^6$Li). Our low-temperature results confirm a
zero-temperature quantum phase transition at a critical polarization
$\sigma_{c0}\approx 0.36$.

This work required the introduction of several novel techniques. A
tomographic reconstruction of local Fermi temperatures and spin
polarization allowed us to obtain the phase diagram for the homogeneous
system, no longer affected by the inhomogeneous density of the trapped
samples. Furthermore, absolute temperatures were obtained using in situ
thermometry applied to the non-interacting fully-polarized Fermi gas in
the outer part of the trapped samples, an ideal thermometer with
exactly-known thermal properties. In contrast to previous
works~\cite{LCJ07,ZSS06b}, this is a direct measurement without any
approximations.

\begin{figure}
\begin{center}
\includegraphics[width=2.5in]{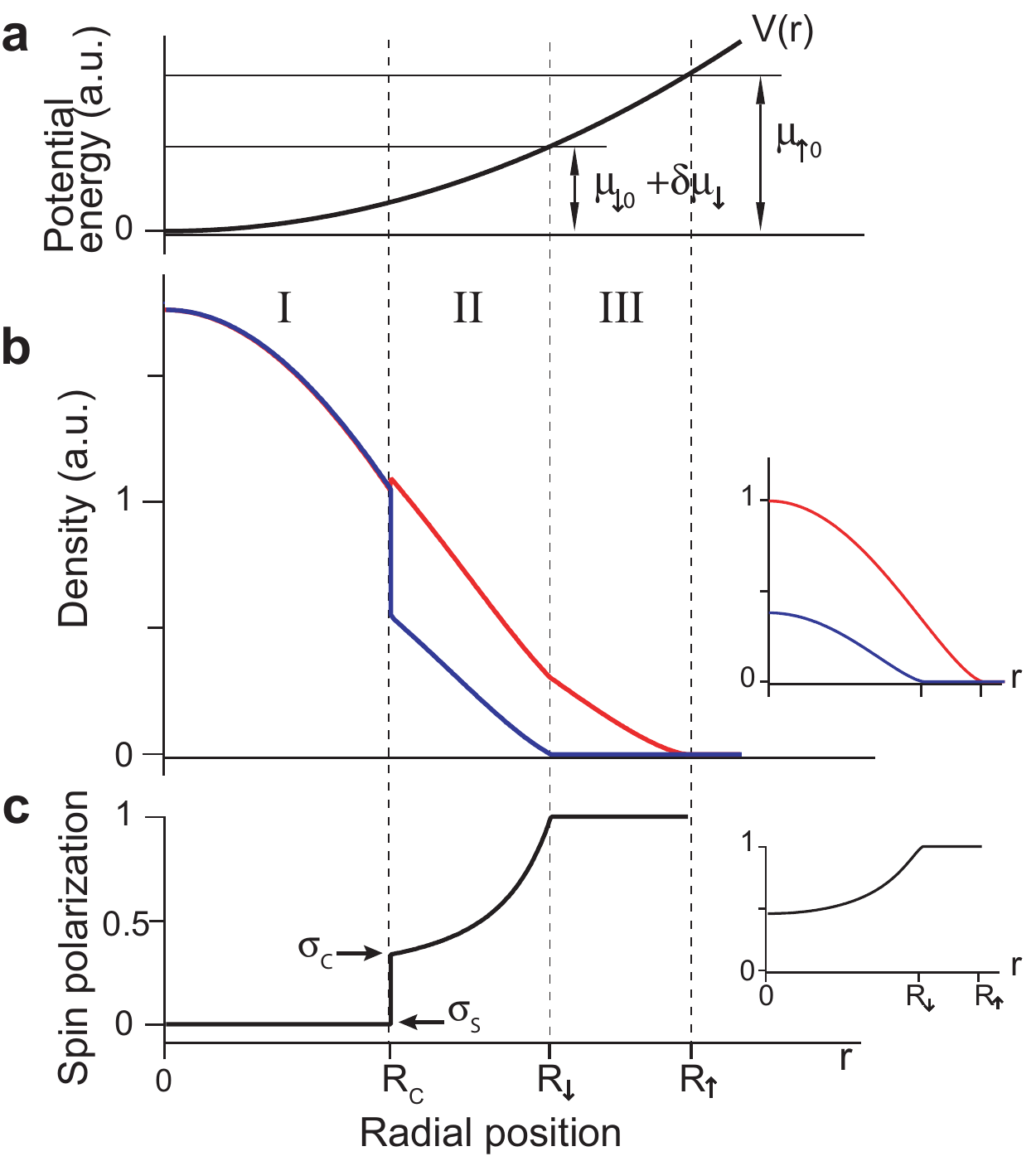}
\caption{Schematic of spatial structure of a strongly interacting Fermi
gas in a harmonic trap. (a) A two-component (spin $\uparrow$ and
$\downarrow$) Fermi mixture is confined in an external potential
$V(r)\propto r^2$ with the chemical potentials of each spin components
$\mu_{\uparrow0,\downarrow0}$ ($\delta \mu_\downarrow$ is the shift for
the spin $\downarrow$ component due to interactions). (b) Density
distributions of the majority component $n_\uparrow(r)$ (red) and the
minority component $n_\downarrow(r)$ (blue). (c) Spin polarization
$\sigma(r)=(n_\uparrow-n_\downarrow)/(n_\uparrow+n_\downarrow)$. At zero
temperature, the trapped Fermi mixture has radially a three-layer
structure: (I) The core region ($0\leq r <R_c$) of a fully paired
superfluid with equal densities of the two components, (II) the
intermediate region ($R_c< r <R_\downarrow$) of a partially polarized
normal gas, and (III) the outer region ($R_\downarrow< r <R_\uparrow$) of
a fully polarized normal gas. The critical polarization $\sigma_c$
($\sigma_s$) is defined as the minimum (maximum) spin polarization of the
normal (superfluid) region. The non-interacting case is shown in insets.
a.u., arbitrary units.\label{f:model}}
\end{center}
\end{figure}

\vspace{0.13in}\noindent\textbf{Spatial structure of a trapped Fermi
mixture}

\noindent Our experiments are carried out in a trapping potential
$V(\vec{r})$. The local chemical potential of each spin component is
given as
$\mu_{\uparrow,\downarrow}(\vec{r})=\mu_{\uparrow0,\downarrow0}-V(\vec{r})$,
where $\mu_{\uparrow 0,\downarrow 0}$ are the global chemical potentials.
When $\mu_{\uparrow0} \neq \mu_{\downarrow0}$ due to imbalanced
populations, the chemical potential ratio
$\eta(\vec{r})=\mu_\downarrow/\mu_\uparrow$ spatially varies over the
trapped sample and thus, under the local density approximation (LDA) the
trapped inhomogeneous sample is represented by a line in phase diagrams of
the homogeneous system. Figure~\ref{f:model} illustrates the spatial
structure of a strongly interacting Fermi mixture in a harmonic trap.  In
the inner region, where $\eta$ is closer to unity, a superfluid with zero
(small) spin polarization will form at zero (low) temperatures, having a
sharp phase boundary against the partially-polarized normal gas in the
outer region. The spin polarization shows a discontinuity at the phase
boundary $r=R_c$, a signature of the phase separation of a superfluid and
a normal gas~\cite{BCR03}. The critical polarization $\sigma_c =
\lim_{r\rightarrow R^+_c}\sigma(r)$ ($\sigma_s =\lim_{r\rightarrow
R^-_c}\sigma(r)$) represents the minimum (maximum) spin polarization for
a stable normal (superfluid) gas.  At higher temperature, the
discontinuity in the density imbalance disappears. The main result of
this paper is the observation and quantitative analysis of such density
profiles. Since we have no experimental evidence, we are not discussing
the exotic partially-polarized phases~\cite{BF07} which could exist only
in the transition layer between the superfluid core and the normal outer
region.

\begin{figure}
\begin{center}
\includegraphics[width=2.8in]{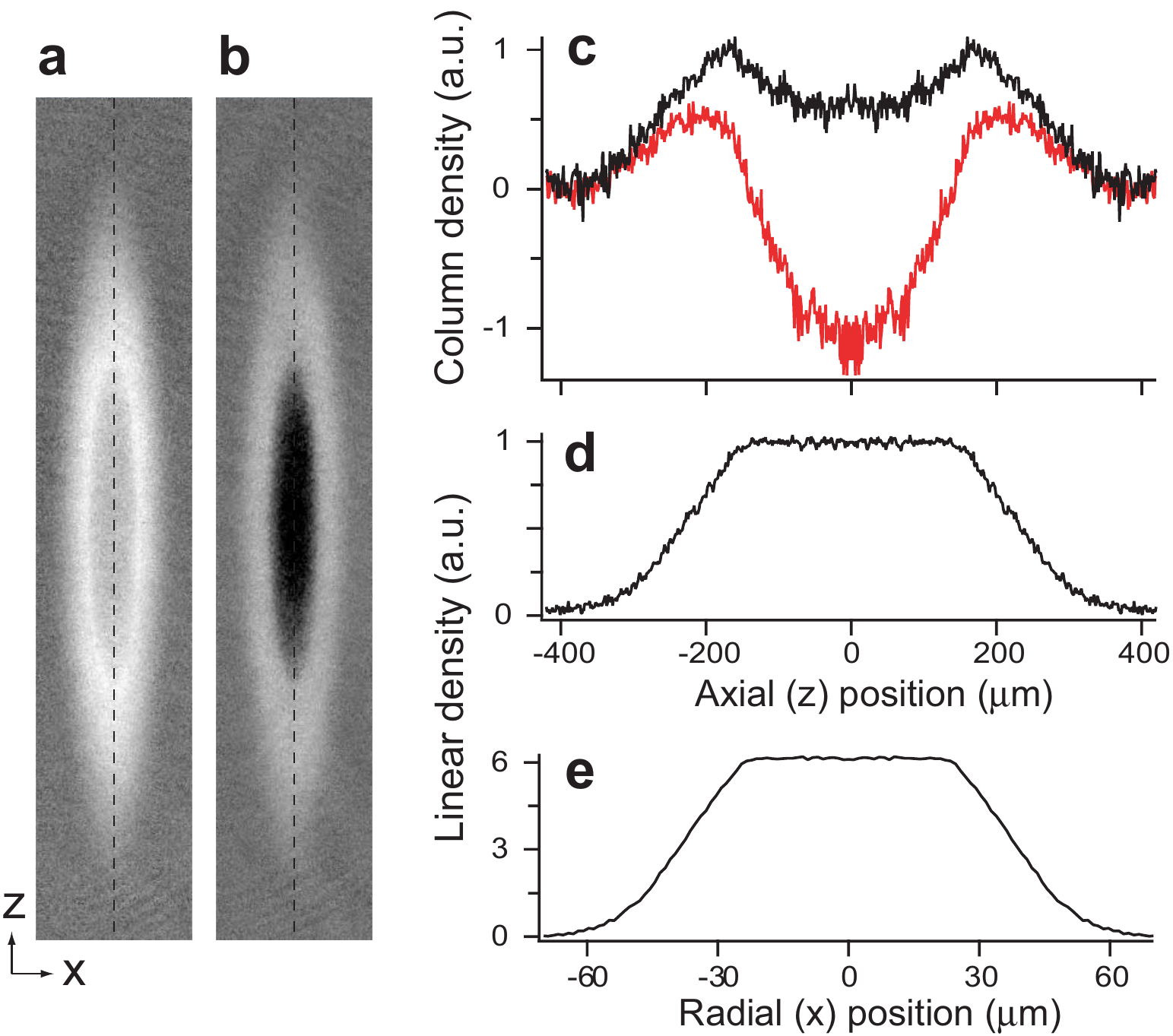}
\caption{Double in situ phase-contrast imaging of a trapped Fermi
mixture. Two phase-contrast images of one sample were taken using
different probe frequencies of the imaging beam, measuring (a) the
density difference $n_{d1} = n_\uparrow-n_\downarrow$ and (b) the
weighted density difference $n_{d2} = 0.76 n_\uparrow - 1.43
n_\downarrow$, respectively. The phase-contrast images show the 2D
distribution of the column density difference,
$\tilde{n}_{d1,2}(x,z)\equiv \int dy~ n_{d1,2}(\vec{r})$ where the
integral describes the line-of-sight integration. The field of view for
each image is $150~\mu$m$\times 820~\mu$m. (c) The distributions of the
column density difference $\tilde{n}_{d1}$ (black) and $\tilde{n}_{d2}$
(red) along the central line (the dashed lines in (a) and (b)). The
profiles of the integrated linear density difference, (d) $\bar{n}_{d1,z}
\equiv \int dx~ \tilde{n}_{d1}(x,z)$ and (e) $\bar{n}_{d1,x} \equiv \int
dz~ \tilde{n}_{d1}(x,z)$ show the identical flattop feature except
scaling. The aspect ratio of the trapping potential was $\lambda=6.15$,
the majority atom number was $N_\uparrow = 5.9(5)\times 10^6$, the
population imbalance was $\delta=44(4)\%$, and the relative temperature
was $T'=T/T_{F0}=0.03(1)$ (see text for definitions).\label{f:doublepc}}
\end{center}
\end{figure}

\begin{figure*}
\begin{center}
\includegraphics[width=4.8in]{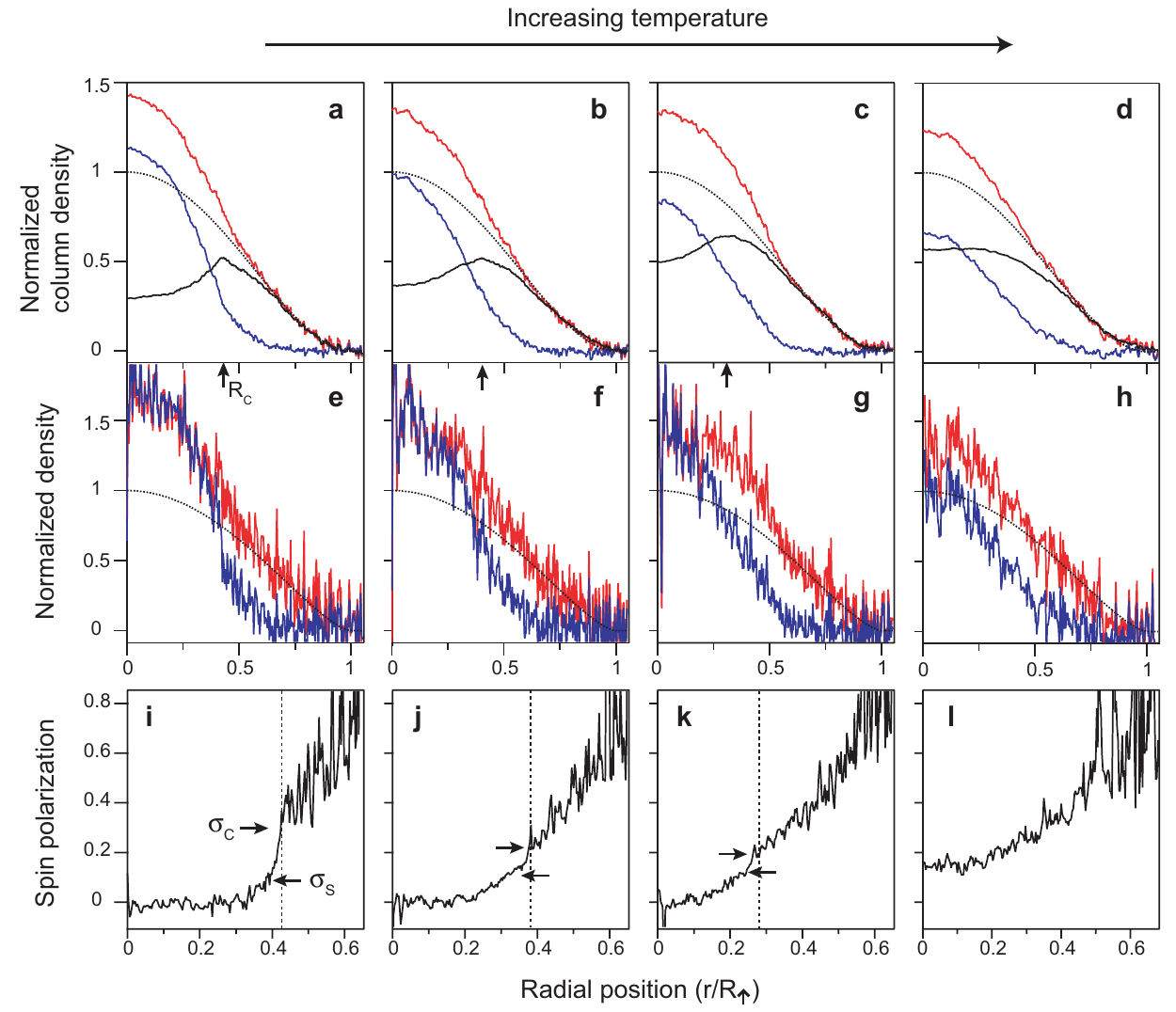}
\caption{Density profiles of trapped Fermi mixtures with imbalanced
populations. The first row (a-d) shows the averaged column density
profiles for various temperatures (red: majority, blue: minority, black:
difference). The majority radius $R_\uparrow$ was determined from the
outer region ($r>R_\downarrow$, $R_\downarrow$: the radius of the minority
cloud) of the majority profiles using a fit to a zero-temperature
Thomas-Fermi (TF) distribution (black dotted lines). The column densities
are normalized by the central value of the fitted TF distribution. The
second row (e-h) and the third row (i-l) show the reconstructed 3D
profiles and the spin polarization profiles $\sigma(r)$ corresponding to
the profiles in a-d. The core radius $R_c$ was determined as the peak
(and/or kink) position in the column density difference (only for a-c),
indicated by the up arrows and the dashed lines. The two spin
polarizations $\sigma_c$ at $r=R_c$ and $\sigma_s$ at
$r=R_c-0.05R_\uparrow$ are marked by the right and left arrows,
respectively. $T'$, $\sigma_c$, $R_c / R_\uparrow$, $R_\uparrow$ (in
$\mu$m), $N_\uparrow$, $\delta$ (in $\%$) and $\lambda$ were respectively:
(a, e, i) 0.03(1), 0.34, 0.43, 385, $5.9(5)\times 10^6$, 44(4), 6.15; (b,
f, j) 0.05(2), 0.24, 0.39, 416, $1.0(1)\times 10^7$, 48(4), 6.5; (c, g,
k) 0.07(1), 0.21, 0.29, 443, $1.2(2)\times 10^7$, 54(4), 6.5; (d, h, l)
0.10(1), not determined, not determined ($\sigma_{r=0}=0.15$ and
condensate fraction~$= 2(1)\%$), 398, $5.3(4)\times 10^6$, 54(4),
7.7.\label{f:profiles}}
\end{center}
\end{figure*}

We prepared a variable spin mixture of the two lowest hyperfine states of
$^6$Li atoms, labeled as $|\uparrow\rangle$ and $|\downarrow\rangle$, at
a magnetic field of 833~G. A broad Feshbach resonance is located at 834~G
and the interactions between the two spin states are resonantly enhanced.
Our sample was confined in a 3D harmonic trap with cylindrical symmetry.
The in situ density distributions of the majority (spin $\uparrow$) and
minority (spin $\downarrow$) components were determined using a
phase-contrast imaging technique. Since the trapped sample was observed
to have an elliptical shell structure of the same aspect ratio
$\lambda=f_\rho /f_z$ as the trapping potential over our entire
temperature range, where $f_\rho$ ($f_z$) is the oscillation frequency in
the transverse (axial) direction (Fig.~\ref{f:doublepc}), we obtained the
low-noise profiles $\tilde{n}(r)$ by averaging the column density
distribution along the equipotenital line (defined as $\lambda^2 x^2+
z^2=r^2$ for a given radial position $r$). The region for averaging was
restricted depending on the type of analysis. The density profiles $n(r)$
were determined from the 3D reconstruction using the inverse Abel
transformation of the column densities $\tilde{n}(r)$~\cite{Bra86}.
Deviations from the trap aspect ratio were only found for the outer
thermal wings and will be discussed below. More detailed description of
the experimental procedure and the image processing is provided in
appendix. Most of our measurements were performed at a total population
imbalance $\delta \approx 50\%$, where $\delta=(N_\uparrow
-N_\downarrow)/(N_\uparrow + N_\downarrow)$ refers to the total atom
numbers in the sample, $N_\uparrow$ and $N_\downarrow$ of the spin
$\uparrow$ and $\downarrow$ components, respectively.

Figure~\ref{f:profiles} displays the radial profiles of the densities
$n_\uparrow(r)$ and $n_\downarrow(r)$ of each component and the
corresponding spin polarization $\sigma(r)$ for various temperatures. The
discontinuity in the spin polarization, clearly shown at very low
temperature, demonstrates the phase separation of the inner superfluid of
low polarization and the outer normal gas of high polarization. At low
temperature the core radius $R_c$ is determined as the kink (and/or peak)
position in the column density difference profile. At high temperature
(but still in the superfluid regime), the discontinuity in $\sigma(r)$
disappears. At our lowest temperature, the radii of the minority cloud
and the core region were measured as $R_\downarrow=0.73(1)R_\uparrow$ and
$R_c=0.430(3) R_\uparrow$ (at $\delta=44(4)\%$), respectively, and agree
with recent theoretical calculations~\cite{LRG06,BF07} within the
experimental uncertainties due to the determination of $\delta$. Here,
$R_\uparrow$ is the radius of the majority cloud, measured as $R_\uparrow
\approx 0.95 R_{TF}$ ($R_{TF}$ is the Thomas-Fermi (TF) radius of a
non-interacting Fermi gas with the same atom number of $N_\uparrow$),
reflecting the attraction between the two components.

\begin{figure}
\begin{center}
\includegraphics[width=2.0in]{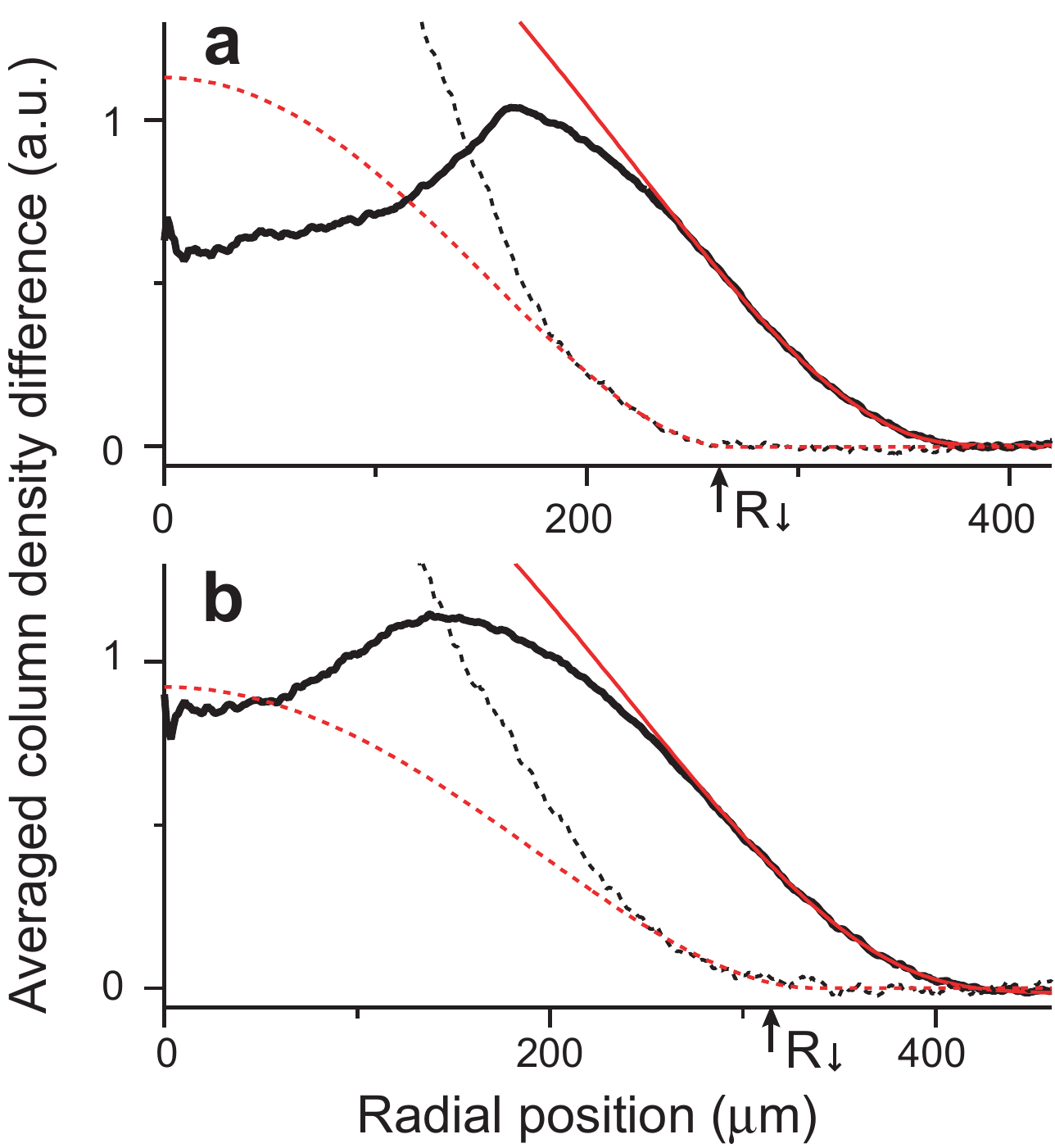}
\caption{Temperature determination using in situ density profiles. The
relative temperature $T'=T/T_{F0}$ (see text for definition) was
determined from the outer region ($r>R_\downarrow$) of the averaged
column density difference profile (black line) fitted to a finite
temperature Fermi-Dirac distribution (red line). The radius of the
minority cloud $R_\downarrow$ was determined from a fit of the wing
profile of the minority component (black dashed line) to a zero
temperature TF distribution (red dashed line). (a) $T'=0.03(1)$ and
$\delta=44(4)\%$, and (b) $T'=0.08(1)$ and
$\delta=46(4)\%$.\label{f:insituT}}
\end{center}
\end{figure}

\vspace{0.13in}\noindent\textbf{Thermometry of a strongly interacting
Fermi gas}

\noindent In most cold atom experiments, temperature is determined from
the density distribution after ballistic expansion, reflecting the
momentum distribution of thermal atoms in the cloud. In the case of a
population-imbalanced Fermi mixture, the outer part of the majority
component, having no spatial overlap with the minority component, is a
non-interacting Fermi gas in thermal equilibrium with the inner part and
fulfills the definition of an ideal thermometer, namely a substance with
exactly-understood properties in contact with the sample to be
characterized. This concept was first introduced in ref.~\cite{ZSS06b},
where temperature was determined from the majority wing profile after
expansion. However, we found that during expansion some of the outer
majority atoms experience collisions with the minority atoms in the inner
region, causing a modification of their kinetic energy  (see appendix).
We avoided this problem by analyzing the in situ profiles. The outer part
of the averaged column density difference profile ($r>R_\downarrow$) was
fit to a finite temperature Fermi-Dirac distribution in a harmonic trap
(Fig~\ref{f:insituT}) and the relative temperature $T'\equiv T/T_{F0}$
was determined, where $k_B T_{F0} = \hbar^2 (6 \pi^2 n_0)^{2/3} / 2m$ is
the Fermi energy of the non-interacting Fermi gas which has the same
density distribution in the outer region as the majority cloud ($n_0$ is
the central density of the non-interacting Fermi gas at zero
temperature).  We verified that anharmonicity of the trapping potential
does not affect the fitted temperature (see appendix).

\vspace{0.13in}\noindent\textbf{Phase diagram for a homogeneous system}

\noindent The critical lines of the phase diagram of a homogeneous
spin-polarized Fermi gas were obtained by determining the local
temperature and spin polarization at the phase boundary. The local
relative temperature $T'_{loc}=T/T_{F\uparrow}$ was derived from the local
density $n_\uparrow(R_c)$ as $T'(R_c)=T/T_{F0} \times (n_0 /
n_\uparrow(R_c))^{2/3}$. Since we observe no jump in the majority density
within our resolution, $T_{F\uparrow}$ is well-defined at the phase
boundary. The critical spin polarizations $\sigma_c$ and $\sigma_s$ were
measured as $\sigma_c=\sigma(R_c)$ and
$\sigma_s=\sigma(R_c-0.05R_\uparrow)$~\footnote{This criterion for
$\sigma_s$ was more robust than a fitting procedure, but precludes that
$\sigma_s$ will be equal to $\sigma_c$ at high temperature.  Therefore,
$\sigma_s$ should be regarded as a lower bound for the polarization of the
superfluid at the phase boundary}. The discontinuity in the spin
polarization profile implies that there is a thermodynamically unstable
window, $\sigma_s < \sigma <\sigma_c$, leading to a first-order
superfluid-to-normal phase transition. As the temperature increases, the
unstable region reduces with  $\sigma_c$ decreasing and $\sigma_s$
increasing.  For high temperature when the bimodal feature in the spin
polarization profile disappears, we recorded the condensate fraction as
an indicator of superfluidity, using the rapid field-ramp
technique~\cite{ZSS06a}. As the temperature decreases, the condensate
fraction gradually increases with a finite central
polarization~\cite{SZS06}.  Such a smooth variation of the density
profile and condensate fraction across the phase transition are
characteristic of a second-order phase transition.

The phase diagram is characterized by the three distinguished points: the
critical temperature $T_{c0}$ for a balanced mixture, the critical spin
polarization $\sigma_{c0}$ of a normal gas at zero temperature, and the
tricritical point $(\sigma_{tc},T_{tc})$ where the nature of the phase
transition changes from second-order to first-order. Due to the lack of a
predicted functional form for the phase transition line in the
$\sigma$-$T$ plane, we apply a linear fit to the measured critical
points, suggesting $T_{c0}/T_{F\uparrow} \approx 0.15$, $\sigma_{c0}
\approx 0.36$ and $(\sigma_{tc}, T_{tc}/T_{F\uparrow}) \approx (0.20,
0.07)$. The value for $\sigma_{c0}$ agrees well with the prediction of
QMC calculation of 0.39~\cite{LRG06}. The extrapolation of the phase
diagram to $\sigma=0$ is only tentative, since the in situ thermometry
could not be applied to small population imbalances due to the narrowness
of the non-interacting outer region. It is possible that the first- and
second-order transition lines meet at the tricritical point with
different slopes, similar to the case of liquid $^3$He-$^4$He
mixtures~\cite{GLR67}.

\begin{figure}
\begin{center}
\includegraphics[width=2.6in]{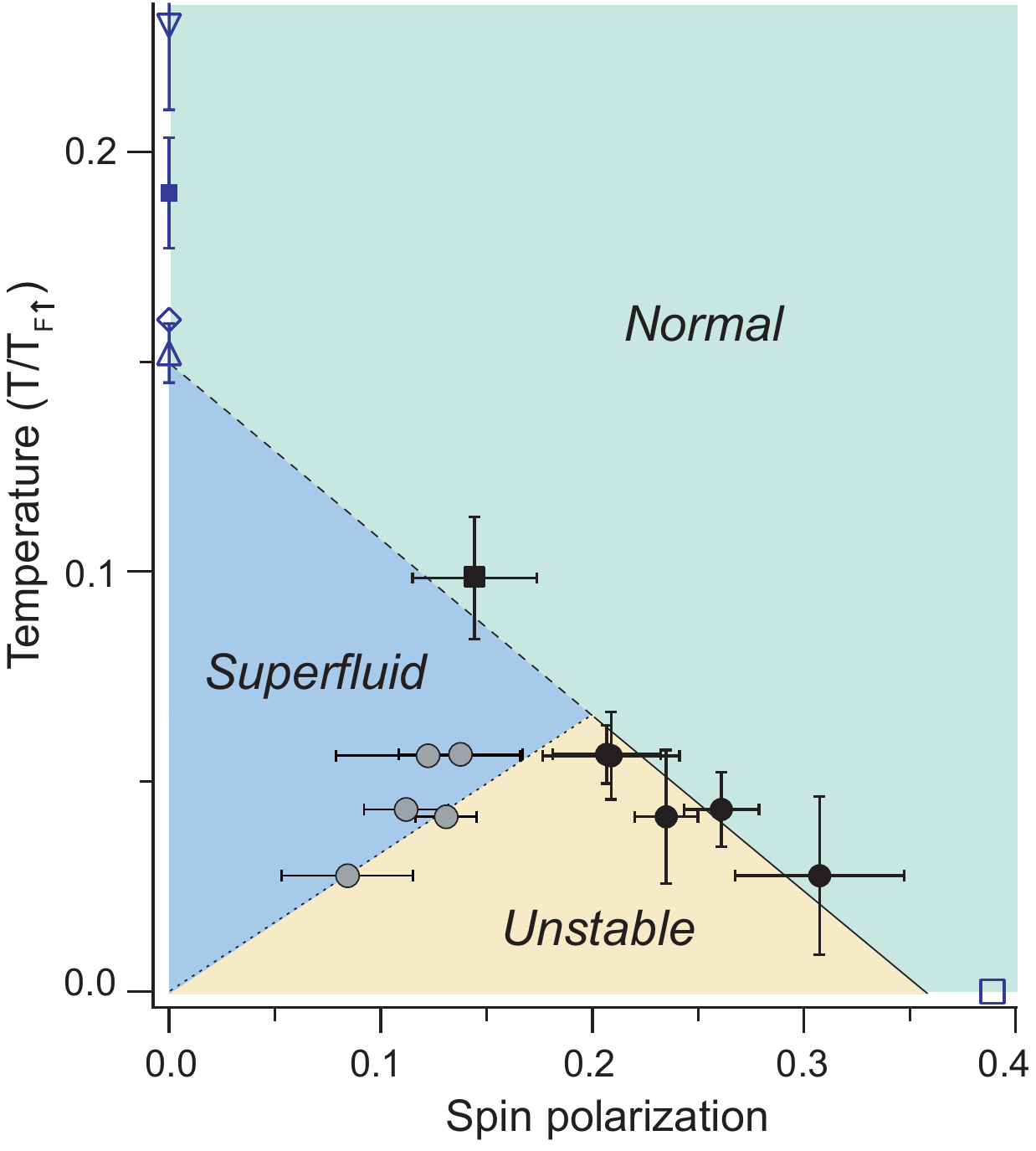}
\caption{$\sigma$-$T$ phase diagram for a homogeneous spin-polarized
Fermi gas with resonant interactions. The critical polarizations
$\sigma_c$ (black solid circles and square) and $\sigma_s$ (gray solid
circles) are displayed along the local $T/T_{F\uparrow}$ at the phase
boundary. The yellow area ($\sigma_s < \sigma < \sigma_c$) represents a
thermodynamically unstable region, leading to the phase separation. Above
the tricritical point, the phase transition in the center of the cloud
was observed by the onset of pair condensation. For this, a cloud was
evaporatively cooled, until it crossed the phase transition on a
trajectory almost perpendicular to the phase transition line (see
appendix). The critical spin polarization and temperature were obtained
by interpolating between points without and with small condensates (black
solid square). The linear fit to the $\sigma_c$'s is shown as a guide to
the eye for the normal-to-superfluid phase transition line. Each data
point consists of five independent measurements and error bars indicate
standard deviation. The blue open symbols show theoretical predictions
for the critical temperature of a homogeneous equal mixture
($\bigtriangledown$: Bulgac \textit{et al.}~\cite{BDM06},
$\bigtriangleup$: Burovski \textit{et al.}~\cite{BPS06}, $\Diamond$:
Haussmann \textit{et al.}~\cite{HRC07}) and the critical polarization at
zero temperature ($\Box$: Lobo \textit{et al.}~\cite{LRG06}). The blue
solid square is the measured critical temperature from Luo \textit{et
al.}~\cite{LCJ07}, multiplied by $\sqrt{\xi}$ with $\xi=0.42$~\cite{CR05}
to obtain local $T/T_F$ at the center. Finite temperature correction may
increase the effective value of $\xi$.\label{f:phasediagram}}
\end{center}
\end{figure}

The zero-temperature phase diagram and in particular the value for the CC
limit are significantly affected by strong interactions in the normal
phase. The CC limit reflects the energetic competition between a
superfluid state and a partially-polarized normal state. When the
chemical potential difference $\delta \mu = \mu_\uparrow -
\mu_\downarrow$ is larger than a critical difference $2 h_c$, the normal
state is energetically favorable and the superfluid state breaks down. In
BCS theory, valid for weak interactions, the critical difference is $h_c
= \Delta/\sqrt{2}$~\cite{CLO62} ($\Delta$ is the pairing gap). At
unitarity, QMC studies predict $h_c = 1.00(5)\Delta(\approx 1.2
\mu)$~\cite{CR05} with the assumption of a non-interacting normal gas.
Here, $\mu = (\mu_\uparrow +\mu_\downarrow)/2$.  The condition
$\mu_{\downarrow c}=\mu-h_c<0$ implies that $n_\downarrow=0$ in a
non-interacting normal gas, and consequently $\sigma_{c0} = 100\%$, i.e.
the absence of a partially-polarized normal phase. Mean-field
approaches~\cite{SR06,GRS06,YD06b,CCH07,PML07}, which cannot treat the
interactions in the normal phase in an accurate way, also predict a high
critical imbalance $\sigma_{c0}>90\%$. Strong interactions between the
atoms in the normal phase, however, have been observed through the
compressed shape of the minority cloud~\cite{ZSS06b} and the shift in the
RF excitation spectrum~\cite{SSS07}. The data in
Figure~\ref{f:phasediagram} clearly establish a zero-temperature CC limit
for $\sigma_{c0}$ in the range of $30\%$ to $40\%$.

The density profiles at our lowest temperature provide quantitative
information on the zero-temperature thermodynamics~\cite{Chevy06,BF07}.
At zero temperature, the global chemical potential of a fully-paired
superfluid in the core is given as $\mu_{s0}= \xi \varepsilon_F = \xi
\hbar^2 (6\pi^2 n_{s0})^{2/3}/2m$ where $\varepsilon_F$ is the local
Fermi energy and $n_{s0}$ is the majority (or minority) density at the
center, whereas $\mu_{\uparrow 0} = \hbar^2 (6\pi^2 n_0)^{2/3}/2m$ and
$\mu_{\downarrow 0} = \eta_0 \mu_{\uparrow 0}$. From the thermodynamic
equilibrium condition $\mu_{s0} = (\mu_{\uparrow 0} + \mu_{\downarrow
0})/2$, we obtain the chemical potential ratio as
\begin{eqnarray}
\eta(r)=\frac{\eta_0 - r^2/ R^2_\uparrow}{1 - r^2/ R^2_\uparrow} =
2\frac{\xi (n_{s0}/n_0)^{2/3} -1}{1 - r^2/ R^2_\uparrow}+1.
\end{eqnarray}
In our coldest sample ($\delta \approx 45\%$), the normalized central
density and the radii for the phase boundary and the minority cloud were
measured to be $n_{s0}/n_0=1.72(4), R_c/R_\uparrow=0.430(3)$, and
$R_\downarrow/R_\uparrow=0.728(8)$, respectively, yielding
$\eta_c=\eta(R_c) \approx 0.03$ and $\eta_\downarrow=\eta(R_\downarrow)
\approx -0.69$ with $\xi=0.42$~\cite{CR05}. Furthermore, the critical
difference is given as $h_c/\mu = (1-\eta_c)/(1+\eta_c)=0.95$. Since
theory clearly predicts $\mu<\Delta$~\cite{CR05,HRC07}, we have
$h_c<\Delta$. If $h_c$ were larger than $\Delta$, polarized
quasi-particles would have negative energies and form already at zero
temperature. Therefore, up to our observed value of $h_c$, the
fully-paired superfluid state is stable, and a polarized superfluid
exists only at finite temperature.

The interface between two immiscible fluids involves a surface energy,
leading to at least a small violation of the LDA. However, the observed
sharp interface along the an equipotential line and the flattop structure
of the linear density difference profiles (Fig.~\ref{f:doublepc}d and e)
imply that corrections to the LDA are smaller than the resolution of our
experiment. These observations are inconsistent with the interpretations
given for the experimental results reported in ref.~\cite{PLK06,PLK06b},
where it has been shown that highly-elongated small samples are deformed
by surface tension~\cite{DSM06b,HS07}. The scaling of those surface
effects to our parameters predicted a deviation of the aspect ratio of
the superfluid core of $\approx 15\%$ from the trap aspect
ratio~\cite{HS07}, whereas we observe this deviation to be smaller than
2\%. Note that surface tension would add energy in the phase-separated
superfluid regime and would shift the CC limit to smaller values.
Ref.~\cite{PLK06,PLK06b} concluded that the CC limit should be
$\delta_{c0}>95\%$ which is ruled out by our observations. We are not
aware of any suggested effect which can reconcile the data of
ref.~\cite{PLK06,PLK06b} with our phase diagram for a resonant
superfluid. To indentify this finite size effect and to fully understand
the nature of the normal state~\cite{SSS07} are still open questions for
imbalanced Fermi gases.

\vspace{0.13in}\noindent\textbf{Conclusions}

\noindent We have established the phase diagram of a homogeneous
spin-polarized Fermi gas with resonant interactions in the $\sigma$-$T$
plane.  This includes the identification of a tricritical point where the
critical lines for first-order and second-order phase transitions meet,
and the final confirmation of a zero-temperature quantum phase
transition, the CC limit of superfluidity, for a gas at unitarity.  So
far, predicted exotic superfluid states such as the breached-pair state
in a stronger coupling regime (``BEC side")~\cite{YD06b,ISdM06} and the
FFLO state in a weaker coupling regime (``BCS side")
~\cite{SR06,MMI06,HL06,PML07,KJT06,YY07} have not been observed, but the
novel methods of tomography and thermometry will be important tools in the
search for those states.

We thank M. W. Zwierlein and A. Keshet for a critical reading of the
manuscript. This work was supported by NSF and ONR.

\section{Appendix}

\noindent\textbf{Experimental details}

\noindent The experimental procedure has been described in previous
publications~\cite{ZSS06a,ZSS06b,SZS06}. A degenerate Fermi gas of $^6$Li
atoms was first prepared in an optical trap, using laser cooling and
sympathetic cooling with $^{23}$Na atoms. A variable spin mixture of the
two lowest hyperfine states $|\uparrow\rangle$ and $|\downarrow\rangle$
(corresponding to the $|F=1/2,m_F=1/2\rangle$ and
$|F=1/2,m_F=-1/2\rangle$ states at low magnetic field) was created at a
magnetic field $B=885$~G. The final evaporative cooling by lowering the
trap depth and all measurements were performed at $B=833$~G. The
temperature of the cloud was controlled by the lowest value of the trap
depth in the evaporative cooling process. The axial trap frequency was
$f_z=23$~Hz. The two transverse trap frequencies are equal within less
than $2\%$.

The optical signal in the phase-contrast imaging is proportional to the
net phase shift of the imaging beam passing through a Fermi mixture, i.e.
$c_\uparrow n_\uparrow - c_\downarrow n_\downarrow \propto n_\uparrow /
(\nu-\nu^0_\uparrow) + n_\downarrow / (\nu-\nu^0_\downarrow)$, where
$\nu$ is the probe frequency of the imaging beam, and $\nu^0_\uparrow$
and $\nu^0_\downarrow$ are the resonance frequencies of the optical
transition for the states $|\uparrow\rangle$ and $|\downarrow\rangle$,
respectively. When the probe beam is tuned to the middle of the two
transitions, i.e. $\nu = \nu_0= (\nu^0_\uparrow+\nu^0_\downarrow)/2$, the
optical signal reflects the density difference
$n_d=n_\uparrow-n_\downarrow$ with $c_\uparrow=c_\downarrow$. In our
experiment, two phase-contrast images of the same sample were taken
consecutively with different probe frequencies, $\nu_1$ and $\nu_2$.  The
two images record the density difference $n_{d1} = n_\uparrow
-n_\downarrow$ and the weighted density difference
$n_{d2}=\alpha_\uparrow n_\uparrow - \alpha_\downarrow n_\downarrow$.
$\nu_1$ was determined by zeroing the optical signal with an equal
mixture and $\alpha_{\uparrow,\downarrow}$ was determined by the signal
ratio between the first and the second image for a highly imbalanced
Fermi mixture with $|\delta|> 95\%$ (an almost fully polarized gas).
Finally, we obtained $n_\uparrow = (\alpha_\downarrow n_{d1} -
n_{d2})/(\alpha_\downarrow-\alpha_\uparrow)$ and $n_\downarrow =
(\alpha_\uparrow n_{d1} - n_{d2}) /(\alpha_\downarrow-\alpha_\uparrow)$.
The difference between $\nu_1$ and $\nu_2$ was chosen to lie between 8
and 13~MHz. The time interval between the two images was $10~\mu$s, and
the pulse duration of each probe beam was $15~\mu$s. Because the probe
beam was off-resonant, no heating effect of the first pulse was observed
in the second image.

Low-noise profiles were obtained by averaging the column density
distribution of phase-contrast images along elliptical equipotential
lines ($\lambda^2 x^2 +z^2 =r^2$). For the measurement of the critical
spin polarization, the averaging region was restricted to $|x|< 12~\mu$m
in order to preserve the sharp features at the phase boundary. The
diffraction limit for our imaging system was about $2~\mu$m. For the
determination of local quantities in the profiles, we averaged over $\pm
5~\mu$m around a given position. For temperature determination, the
averaging region was restricted to an axial sector of $\pm 60^\circ$ to
avoid corrections due to radial anharmonicities (see appendix). The
relative temperature $T'$ is determined as $T'\equiv T/T_{F0}=(-6 Li_3
(-\zeta))^{-1/3}$, where $\zeta$ is the fugacity obtained from the fit
($Li_s(z)\equiv \sum_{k=1}^\infty z^k/k^s$ is the Poly-Logarithmic
function of order $s$).

\begin{figure}
\begin{center}
\includegraphics[width=2.5in]{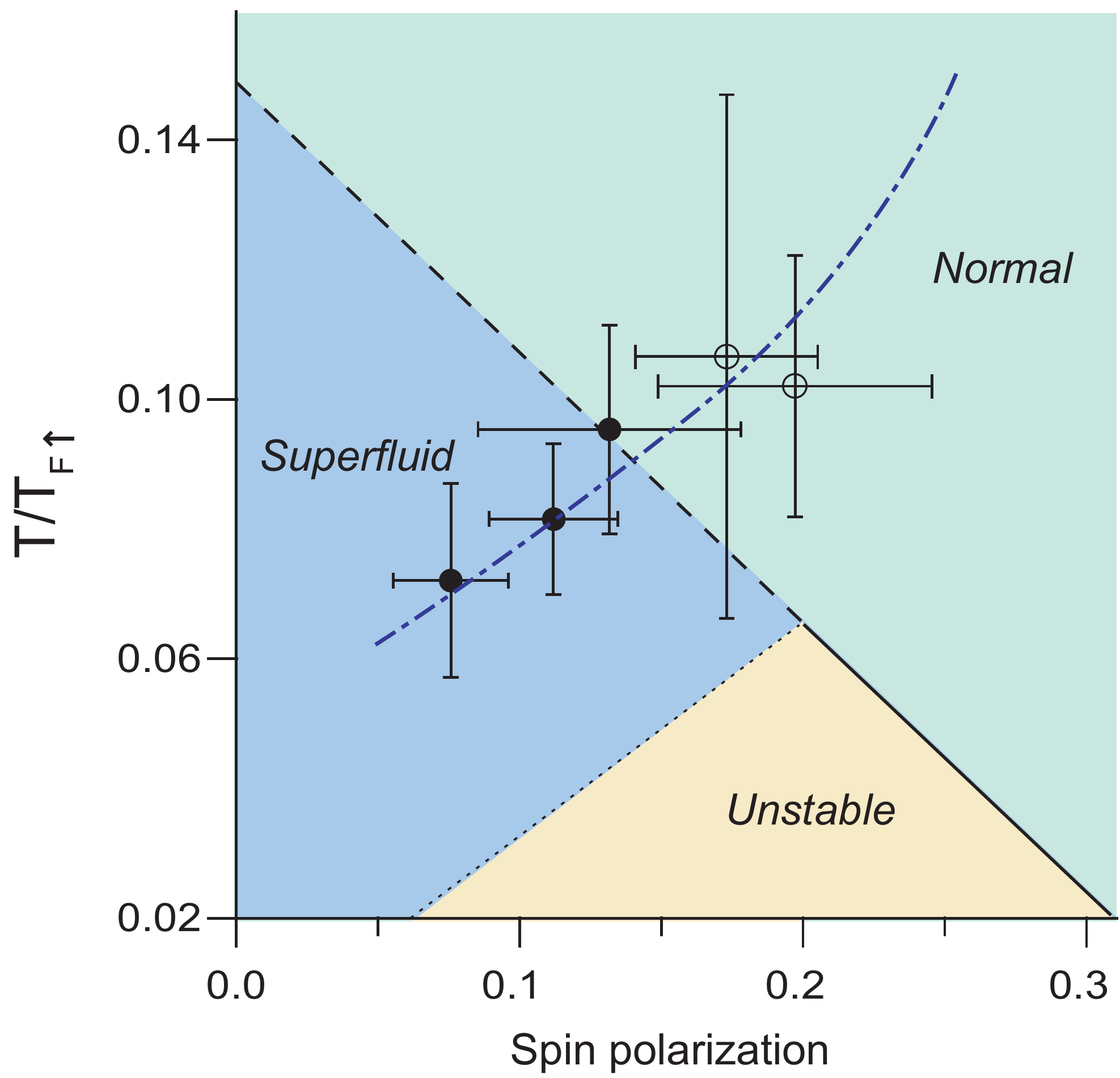}
\caption{Trajectory of the center of a cloud in the phase diagram during
the cooling process. Above the tricritical point, the
normal-to-superfluid phase transition was observed by the onset of pair
condensation in the evaporative cooling process. The local spin
polarization and temperature at the center of the cloud was measured
(black solid (open) circles with (without) condensate fraction) and the
critical point was obtained by linearly interpolating with the condensate
fraction. The dashed-dot line shows a guide line for the trajectory of
the cloud center. The population imbalance of the sample was $\delta
\approx 55\%$. A non-interacting mixture with this imbalance has a spin
polarization $\sigma \approx 30\%$ at the center at zero
temperature.\label{f:pdinset}}
\end{center}
\end{figure}

\vspace{0.13in} \noindent\textbf{Thermometry of ultracold Fermi gases}

\begin{figure}
\begin{center}
\includegraphics[width=3.2in]{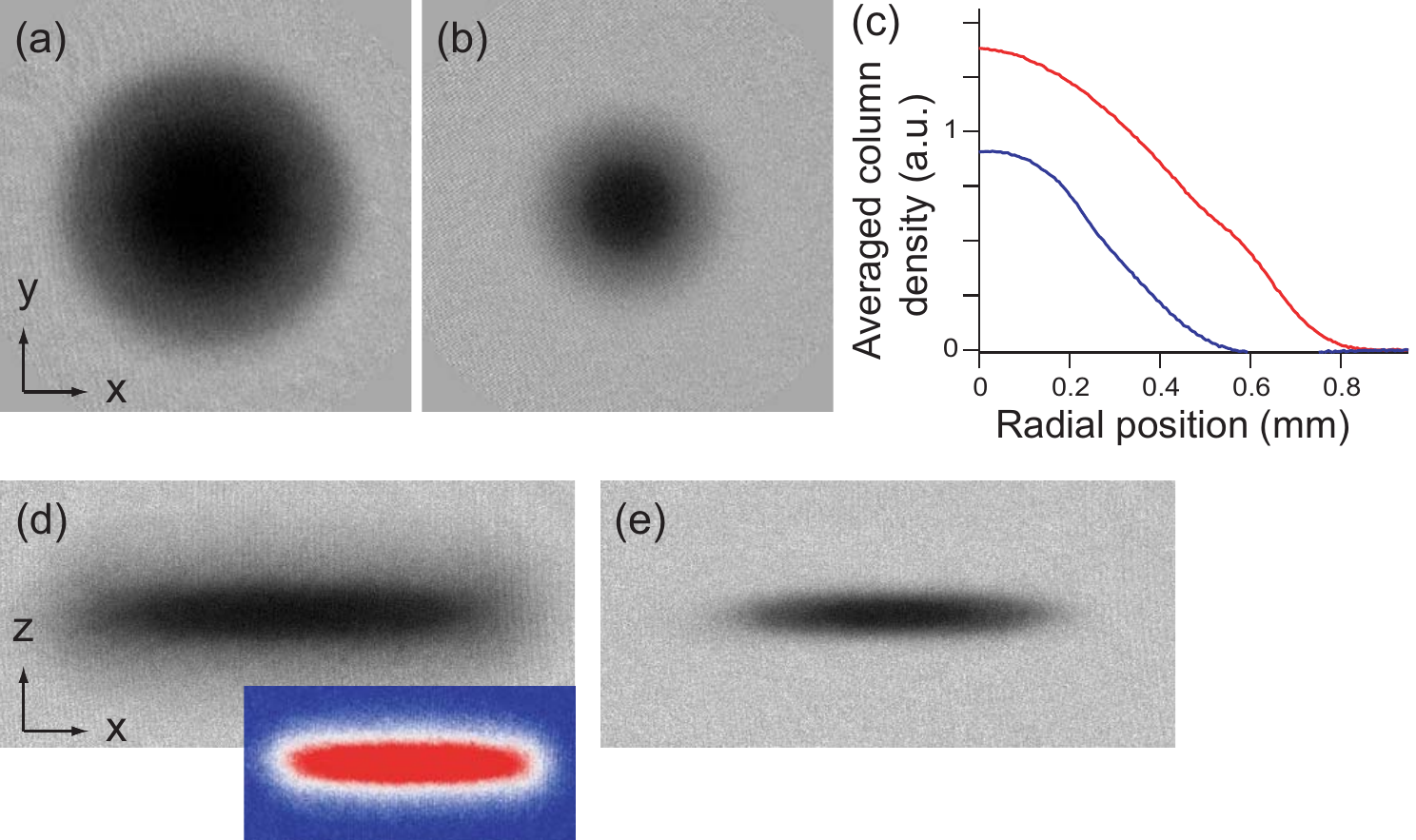}
\caption{Expansion of a population-imbalanced Fermi mixture. The
absorption images of (a, d) the majority and (b, e) minority components
were taken along (a, b) the axial $z$ and (d, e) transverse $y$
directions after expansion. (c) The azimuthally averaged column density
profiles of the majority (red) and the minority (blue) cloud are obtained
from (a) and (b), respectively. The excess majority atoms in the outer
region interact with the core during expansion. The contour lines of the
outer part of the majority cloud (color inset) are not elliptical and
have the shape of a horse-track.  This shows that the minority cloud
pushes the outer majority atoms in the transverse direction, which is also
indicated by the hump of the majority profile at the edge of the minority
cloud. The population imbalance was $\delta \approx 55\%$.
\label{f:expansion}}
\end{center}
\end{figure}

\noindent In our previous work~\cite{ZSS06b,SZS06}, temperatures have
been determined by fitting the spatial wings of the majority component
after expansion.  However, we found that one can neglect collisions with
the minority atoms in the core only for large population imbalances.  In
a simplified picture, one can regard collisions with the inner core as
collisions with a moving wall, which moves outward radially and inward
axially (due to the magnetic trapping potential).  This results in
different average kinetic energies (transversely and axially) of the free
majority atoms in the outer region. Figure~\ref{f:expansion} shows the
density distribution of the majority and minority components after
expansion. Although the temperature has been overestimated by only $20\%$
for typical experimental conditions ($\delta \approx 60\%$) in
refs~\cite{ZSS06b,SZS06}, we do not regard this technique as
well-calibrated absolute thermometry.

One other concept for thermometry determines temperature as the
derivative of entropy with energy. So far, this concept could be
implemented only for balanced fermion mixtures with certain
approximations, and due to the need of determining a derivative, could
only be used to obtain temperatures averaged over a certain
range~\cite{LCJ07}.

\vspace{0.13in} \noindent\textbf{Anharmonicity of the trapping potential}

\noindent For the determination of temperatures from the spatial in situ
profiles it was necessary to address the anharmonicity of the trapping
potential. Our trap is generated by a weakly focused (beam waist $w
\approx 125~\mu$m) infrared Gaussian laser beam (wavelength 1064~nm) near
the saddle point of a magnetic potential. The total trapping potential is
given as
\begin{eqnarray}
V(\rho,z)=U_0 \exp(-\frac{2 \rho^2}{w^2})+ \frac{m (2\pi f_z)^2}{2}
(-\frac{\rho^2}{2}+z^2),
\end{eqnarray}
where $\rho^2= x^2+y^2$. Note that gravity has been compensated by a
magnetic field gradient.  The axial confinement comes mainly from the
magnetic potential with oscillation frequency of $f_z=23$~Hz. The
transverse magnetic potential is anti-trapping and limits the trap depth
as
\begin{eqnarray}
U=\frac{1}{4} m (2\pi f_\rho)^2 w^2 ( 1- \frac{f^2_z}{2 f^2_\rho} \ln (
\frac{2 f^2_\rho + f^2_z}{f^2_z} ) ),
\end{eqnarray}
where $f_\rho$ is the transverse oscillation frequency in the central
harmonic region. When the trap depth is comparable to the Fermi energy of
a sample, the transverse anharmonicity will affect the shape of the
cloud. Although in our experiments, the inner core and the outer cloud
had the same aspect ratio as the trapping potential, indicating the
absence of anharmonic effects, anharmonicities were not negligible in the
spatial wings used to determine the temperature.

\begin{figure}
\begin{center}
\includegraphics[width=2.4in]{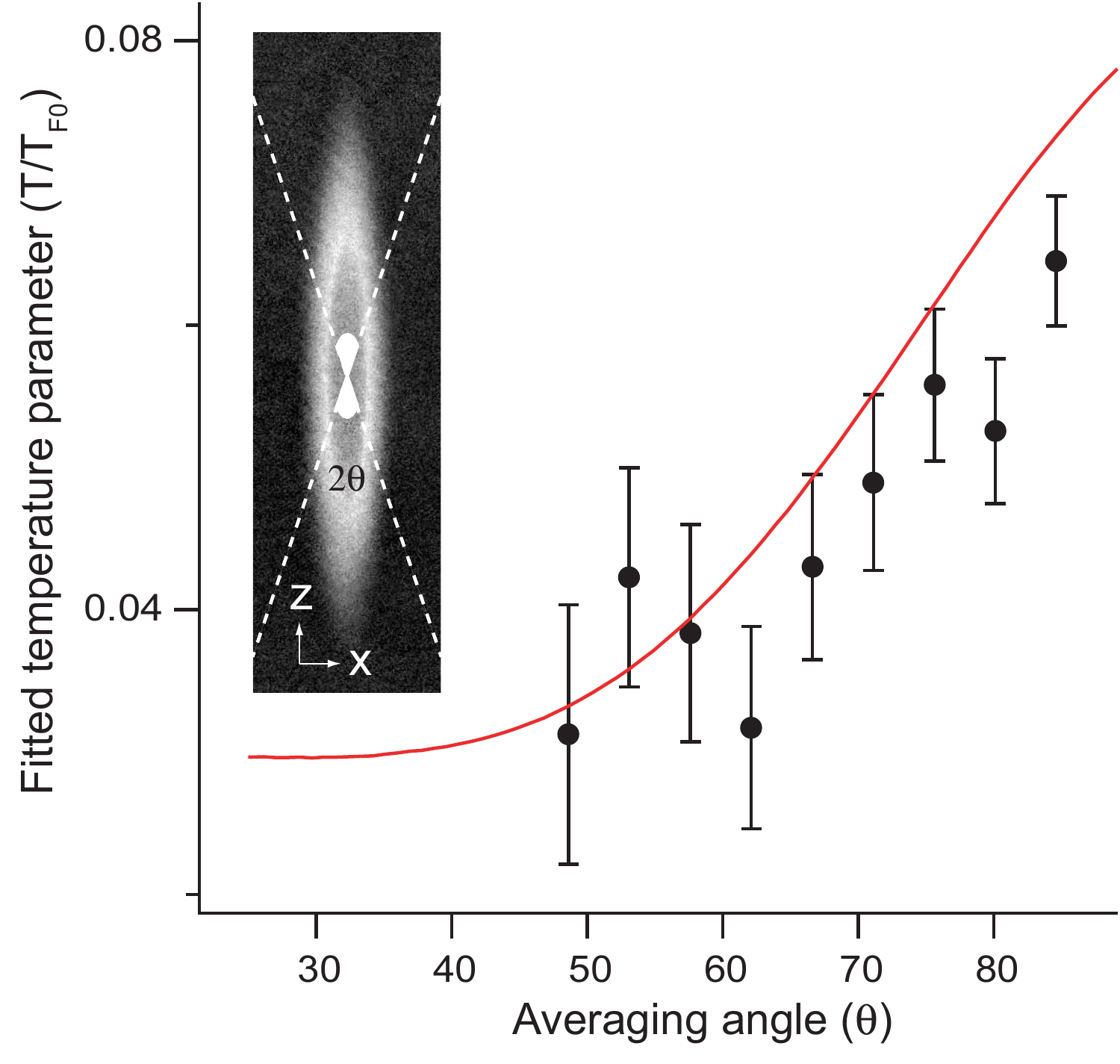}
\caption{The temperature of the cloud was determined for various angles
$\theta$ of the averaging sector. For a large angle, the large-$x$ region
is included in the averaged profile, resulting in a broadening of the
spatial wings and consequently higher value of the fitted temperature.
The red line shows the results of a simulation using the same parameters
as the experiment ($\lambda=f_\rho/f_z=6.15$, $T_{F0}=1~\mu$K and the
trap depth $U/k_B = 2~\mu$K).\label{f:anharmonicity}}
\end{center}
\end{figure}

This issue was addressed by adjusting the angular averaging region
(Fig.~\ref{f:anharmonicity}). Since the trapping potential is only
anharmonic for large $\rho$, we could reduce the effect by decreasing the
angle of the averaging sector around the $z$ direction. Both the
experimental data and an exact simulation for an ideal Fermi gas show
that the fitted temperature remains almost constant up to a certain angle
and then increases when the averaging sector includes more of the
transversely outer region. In our temperature determination, we chose the
averaging sector to be $\pm 60^\circ$ which was large enough to create
low-noise profiles, but kept the effect of the anharmonicities to below
$10\%$. The 1D fit to angularly averaged profiles was computationally
more efficient than a 2D fit to a selected region of the image. In a 2D
fit, one could also include anharmonic terms in the fitting function.

\vspace{0.13in} \noindent\textbf{Polarized superfluid at finite
temperature}

\noindent When the two spin components have a chemical potential
difference $2h$, the BCS-type superfluid has two branches of
quasiparticles with excitation energies $\sqrt{(\epsilon_k- \mu)^2
+\Delta^2} \pm h $ where $\epsilon_k=\hbar^2 k^2 / 2 m$. At finite
temperature, the superfluid is polarized due to the large thermal
population of the lower branch compared to the upper branch. An
interesting situation arises when $h$ becomes larger than $\Delta$, i.e.
the lower branch has negative energy quasiparticles, implying that even
at zero temperature the superfluid state would have a finite
polarization. Our experiments show $h_c < \Delta$ at very low temperature,
suggesting that a polarized superfluid state exists only at finite
temperature. The breached-pair state with $h_c>\Delta$ at zero
temperature has been predicted in a stronger coupling region (on the BEC
side of the Feshbach resonance). Since $\Delta$ gradually decreases with
higher temperature, it might be possible to have $h_c>\Delta$ at finite
temperature, at least in the weakly-interacting BCS limit where $\Delta$
smoothly approaches zero at a second order phase transition point. One
interesting problem is identifying this gapless region of $h>\Delta$ in
the phase diagram for various coupling regimes.

\end{document}